# The Discovery of "Tatooine": Kepler-16b

Laurance R. Doyle

Institute for the Metaphysics of Physics

Principia College

Elsah, Illinois 62028

And

SETI Institute

189 Bernardo Avenue

Mountain View, CA 94043

Abstract

We describe the discovery of Kepler-16b, the first widely accepted detection of a circumbinary planet.

The discovery of Kepler-16b was the first fully charcterized (mass, radius, and orbit determined) detection of a circumbinary planet (CBP) -- a planet that circles two stars (a review of CBP detection can be found in Doyle and Deeg, 2018). Initial evidence of it was found by the author, and his assistant/colleague (Dr. Robert Slawson) in the light curve of the Kepler eclipsing binary (EB) star system KIC-12644769 in March of 2011, and the discovery paper was accepted in August of the same year. Before its discovery, there was a dichotomy of opinions as to whether such CBPs could exist, with the general consensus on one side being that the central binary may dissipate the protoplanetary disc so that no planets could form, (or excite planetesimal relative velocities so that in-situ formation near the binary is difficult; see Scholl et al. 2007), and on the other side that the gravitational interaction of the binary with the protoplanetary disc might encourage planet formation via, for example, the agitation produced by density waves. It should be noted that others recognized that gas giants planets, once formed, could migrate towards the stars and stop before going unstable (e.g., Nelson 2003).

Evidence of CBPs with large orbital periods had been inferred previously for several systems by variations in the timing of the central pulsar or the binary eclipse periodicities. But Kepler-16b was the first planet whose binary stars were close enough be seen as solar-sized discs from its sky (and hence the nickname "Tatooine" after the double sunset seen from the fictional planet Tatooine in the movie "Star Wars"). As the two stars of the Kepler-16AB system have quite different luminosities, their surface brightnesses -- and therefore the depth of their planetary transits -- were quite different. The primary transit across Kepler-16A, a K-dwarf star, produced a 1.7% drop in brightness, while the secondary transit across Kepler-16B, an M-dwarf star, produced an 0.01% drop in the brightness of the light curve. During Kepler spacecraft observations the sequence of Kepler-16AB planetary transits recorded were in the order: primary transit, followed by a secondary transit, then a secondary transit, followed by a primary transit, then a primary transit again, followed by a secondary transit. This was unequivocal proof of a bound third body in a circumbinary orbit around the two stars Kepler-16AB. The alternative explanation, a background false positive of a grazing EB system, could not have produced such a reversal in the order of the light curve events. This would have required a background EB to stop and reverse its mutual orbital direction, which would not have been astrophyically possible.

From the depths of the transits and the known sizes of the two stars -- measured spectroscopically from ground-based observations -- we knew that the transiting object could indeed be planetary in size because of the planet-to-stars area ratio. If one takes the smallest main-sequence star – excluding M-dwarfs for now – with the largest possible planet transiting it, an upper limit of about 6% drop in the brightness of the light curve is the result. However, in addition, the areas of both EB component stars must be divided into the CBP planetary disc area to obtain the approximate transit depth (assuming for now, equal stellar surface brightness flux per unit area). In order to ascertain if the transiting object was indeed a circumbinary object of planetary mass we had to look at the EB timing residuals – the observed minus calculated (O-C) times of the stellar eclipses. This implies a massive third object (e.g., a CBP) encircling the eclipsing binary (EB) system since the latter will be offset around an EB-CBP barycenter as the planet orbits around it, causing the timing of the eclipses to drift periodically (although the EB period is essentially unchanged). This is called the light (travel) time effect, or "LTE" since the offset of the EB system will cause a change in the times the mutual stellar eclipses are received on Earth. Due to this offset, the stellar eclipses occur on time, late, on time, early, on time, etc. in a periodic manner, where the total amplitude of the time change is twice the light travel time across the distance from the EB-CBP barycenter to the (center of the) EB system.

The O-C residuals, for the LTE effect, were of a low enough amplitude to constrain the circumbinary mass to be that of about the least massive brown dwarfs. This offset is, of course, a projected effect so that Msini is what is detected, (where M is the mass of the circumbinary object and i is its orbital inclination.) But since the circumbinary object transited (or grazed) the stars, the inclination was known to be very close to 90 degrees – edge-on to our line-of-sight from Earth – so that $\sin i \sim 1$.

However, further constraints could be placed upon the mass of the circumbinary object using these same O-C measurements by a dynamical effect originally developed by Z. Kopal in the 1950s and reapplied in our case by one of the coauthors of the discovery paper (Daniel Fabrycky; see Borkovits et al. 2003). This has been called the "dynamical effect" and is a direct gravitational tug of the circumbinary object upon the individual stellar components of the binary system (as opposed to a general offset of the whole EB system about the EB-CBP barycenter). The dynamical effect plays an increasing role in the changes of the O-C timings with the circumbinary object's orbital period decrease. This works in the opposite direction compared to the increasing amplitude of the LTE affect with increasing circumbinary object orbital period. This gravitational "tug" will cause a predictable periodic change in the individual stellar eclipse times. As stated, this method increases in sensitivity the shorter the circumbinary orbital period. These dynamical constraints were thus applied to the O-C residuals of the Kepler-16AB eclipses and it was found that the circumbinary object indeed had a planetary mass – about the mass of the planet Saturn.

We thus had the first direct detection of a CBP (the detection of the reflected or deflected light, or the detection of the blocked light – the shadow -- of a planet, being considered here as direct detection methods, as opposed to planets detected by their effect on their parent stars like, e,g,, radial velocity variations).

In an early scene from the science fiction movie Star Wars – produced and directed by George Lucas -- the hero Luke Skywalker watches a double sunset from the hypothetical circumbinary planet Tatooine. Since the discs of the two stellar components of the Kepler-16AB-b system, as seen from the distance of the planet, would appear as discs of comparable size to that of the Sun as seen from the Earth, the nickname "Tatooine" for the planet suggested itself. Upon acceptance of the discovery paper, the author sent around a congratulations to the team along with a note that we should ask George Lucas if we could nickname the planet "Tatooine." One of the NASA project managers of the Kepler Spacecraft actually called George Lucas and asked if he could be at the press conference. He could not attend but sent the Chief Creative Officer of his graphics company – Industrial Light and Magic -- Dr. John Knoll. Dr. Knoll said, "When [George Lucas] envisioned Tatooine, he was using a visual shorthand to wordlessly show that we're not on Earth and we're in this exotic place." Knoll (2011).

Thus the nickname "Tatooine" stuck for the Kepler-16b CBP.

Although the gravitational lensing method for extrasolar planet detection can detect Earth-mass planets, a follow-up study of such systems is not generally possible as such gravitational lensing alignment events do not repeat. Therefore, in general, if one would like to detect Earth-sized planets with current technology the transit method is the most currently viable method. This can be understood when one examines what is being detected. For example, if one is using radial velocity variations in a single solar-mass star to detect a jovian-mass planet, one would generally be comparing the mass ratio, which would be about 1000: 1. For planet detection by transit one is comparing the area ratio which is about 100: 1, in principle a much easier detection.

The Kepler-16AB-b system was of particular interest because the primary star – Kepler-16A – was slowly rotating yet it produced quite a large amount of starspot activity, a somewhat anomalous behavior. At the time the smaller stellar component –

Kepler-16B – was the smallest M-star yet measured by direct eclipse and maintained the general attribute of being 10% larger than current models of the small end of the Main Sequence predicted.

Having been detected by double transit, the orbital plane of Kepler-16b is very close to the mutual orbital planet of the two stellar components of the system, and the rotation axis of Kepler-16A was also found (via the Rossiter-McLaughlin effect) to be perpendicular to these orbital planes (within error bars of about 18 degrees). Thus this system could provide an interesting study of the possible angular momentum formation or evolution of such systems.

An interesting aspect of CBP systems is that the orbital nodes of the planet precess across the line-of-sight of the Earth so that transits come and go. In the case of Kepler-16b, the transits have currently become unobservable due to this precession process. The transits across the secondary star precessed out of the line of sight in 2014 and will not appear again for over four decades. Transits across the primary star continued until early 2018 and have now disappeared for about 24 years. It will be fun if astronomers from the next generation decide to look for these transits to reappear. One is struck by the fact that such CBP transits are only detectable from an Earthlike orbital perspective for less than typically half of their precession lifetimes.

Since the discovery of Kepler-16b about a dozen CBPs have been discovered to date. CBPs are not rare, and there must be at least tens of millions of them in the Milky Way Galaxy (Welsh et al. 2012). Thus Kepler-16AB-b provided the first universally accepted confirmation that such systems existed and unequivocal evidence that the sunset scene from Star Wars might not be so exotic after all.